**Comet 41P/Tuttle-Giacobini-Kresak, 45P/Honda-Mrkos-Pajdusakova, and 46P/Wirtanen:**
**Water Production Activity over 21 Years with SOHO/SWAN**

Short Title: SOHO/SWAN Observations of 41P, 45P and 46P

M.R. Combi[1], T. Mäkinen[2], J.-L. Bertaux[3], E. Quémerais[3], S. Ferron[4]
and R. Coronel[1]

[1]Dept. of Climate and Space Sciences and Engineering
University of Michigan
2455 Hayward Street
Ann Arbor, MI 48109-2143
*Corresponding author: mcombi@umich.edu

[2]Finnish Meteorological Institute, Box 503
SF-00101 Helsinki, FINLAND

LATMOS/IPSL
Université de Versailles Saint-Quentin
11, Boulevard d'Alembert, 78280, Guyancourt, FRANCE

ACRI-st, Sophia-Antipolis, FRANCE

ABSTRACT

In 2017, 2018, and 2019, comets 46P/Wirtanen, 45P/Honda-Mrkos-Pajdusakova, and 41P/Tuttle-Giacobini-Kresak all had perihelion passages. Their hydrogen comae were observed by the Solar Wind ANisotropies (SWAN) all-sky hydrogen Lyman-alpha camera on the SOlar and Heliospheric Observer (SOHO) satellite: comet 46P for the fourth time and comets 45P and 41P for the third time each since 1997. Comet 46P/Wirtanen is one of a small class of so-called hyperactive comets whose gas production rates belie their small size. This comet was the original target comet of the Rosetta mission. The Solar Wind ANisotropies (SWAN) all-sky hydrogen Lyman-alpha camera on the SOlar and Heliospheric Observer (SOHO) satellite observed the hydrogen coma of comet 46P/Wirtanen during the apparitions of 1997, 2002, 2008, and 2018. Over the 22 years, the activity decreased and its variation with heliocentric distance has changed markedly in a way very similar to that of another hyperactive comet, 103P/Hartley 2. Comet 45P/Honda-Mrkos-Pajdusakova was observed by SWAN during its perihelion apparitions of 2001, 2011, and 2017. Over this time period the activity level has remained remarkably similar, with no long-term fading or abrupt decreases. Comet 41P/ Tuttle-Giacobini-Kresak was observed by SWAN in its perihelion apparitions of 2001, 2006, and 2017 and has decreased in activity markedly over the same time period. In 1973 it was known for large outbursts, which continued during the 2001 (2 outbursts) and 2006 (1 outburst) apparitions. However, over the 2001 to 2017 time period covered by the SOHO/SWAN observations the water production rates have greatly decreased by factors of 10—30 over corresponding times during its orbit.

Key Words: Comets; Cometary Atmospheres; Comet 41P/Tuttle-Giacobini-Kresak, 45P/Honda-Mrkos-Pajdusakova, Comet 46P/Wirtanen

1. Introduction

Comet 46P/Wirtanen (hereafter 46P) was discovered in 1948 by Carl Wirtanen (Klemola 1991). It is a Jupiter Family Comet (JFC) with a current orbital period of 5.4 years and a perihelion distance of 1.05 AU. Comet 46P was the original target comet of the European Space Agency Rosetta mission that was changed to 67P/Churyumov-Gerasimenko after the initial planned launch was delayed because of concern over the launch vehicle. Comet 46P has been observed on a number of apparitions since its discovery, especially the last few. The 2008 apparition was particularly poor as the comet was too close to the Sun in the sky. In 2018, however, the comet made an extremely close pass to the Earth (~0.1 AU) and was therefore of great interest to observers.

Comet 46P is one of a small class of so-called hyperactive comets whose water production rate, as well as overall activity, is large compared to its small size, having a radius of ~0.6 km (Lamy et al. 2004), making it comparable in size and activity to another hyperactive comet, 103P/Hartley 2 (A'Hearn et al. 2011). There was some effort to observe comet 46P/Wirtanen in 1997 when it was selected as the original target comet for the European Space Agency mission Rosetta. Various water production rate determinations from the 1997 apparition were summarized by Fink and Combi (2004), who also reanalyzed published observations of water dissociations products such as H, OH, and OI, as well as the other common ground-based species.

Comet 45P/Honda-Mrkos-Pajdusakova (hereafter 45P) was discovered by Minoru Honda, Antonín Mrkos, and Ludmila Pajdušáková on 1948 December 3. It is a JFC with a perihelion distance of only 0.53 AU, so it is quite active at perihelion despite being a relatively small object with an effective diameter of only ~0.8 km. See Combi et al. (2019) and Moulane et al. (2018) for summaries of measurements of the nucleus size. Comet 45P is much less active than 46P when both are near a heliocentric distance of about 1 AU.

According to Fink (2009), 45P is in the Tempel 1 type compositional classification with low $C_2$ and normal $NH_2$ abundances compared to CN and $H_2O$. Infrared spectroscopic measurements (DiSanti et al. 2017) indicate that CO was depleted in 45P compared with the median value of 10 Cloud comets (OCCs), but the relative abundances of other volatile species ($CH_4$, $C_2H_6$, $C_2H_2$, $H_2CO$, and $NH_3$) places it at the low end to the middle of median OCCs and at the higher end of JFCs, though the statistical sample of JFCs is still rather small. The abundance of CH3OH, on the other hand, was rather high, even compared with median OCCs. More recently Dello Russo et al. (2020) have published IR observations of 45P taken a month later than the near perihelion observations (~0.55 AU) of DiSanti et al. when the comet was at ~1 AU from the Sun but only 0.08 AU from the Earth. There were significant changes in the relative abundances of $C_2H_6$, $H_2CO$, HCN and $CH_3OH$ compared to $H_2O$ from those determined by DiSanti et al. (2017) by factors of 1.5 to 3. Whether the differences are due to actual changes in production rates with heliocentric distance, a role for an icy grain source for some species, or to the much smaller spatial scale of the Dello Russo et al. observations enabled by the very small geocentric distance, remains to be seen.

Comet 41P/Tuttle-Giacobini-Kresak (hereafter 41P) is also a JFC with an orbital period of 5.4 years and a perihelion distance of 1.05 AU. Comet 41P is known for outbursts. There were two ~8 mag outbursts in 1973 reported by Kresak (1974), 6 and 8 mag outbursts in 1995, and 6 and 5 mag outbursts in 2001 (Yoshida http://www.aerith.net/comet/catalog/0041P). There were no reported outbursts in 2006 or 2017. It was not observed in 2011 due to poor geometry. Also interesting in 2017 is the reported decrease in the rotation rate of the nucleus by Bodewits et al.

(2018) and Schleicher et al. (2019) with an increase of the rotation period from 20 to more than 50 hours in about 2 months from early March to early May 2017. Moulane et al. (2018) place the $C_2$/CN ratio in 41P into the range of the typical class of comets, however the $C_2$/OH and CN/OH ratios, while still within the typical class range, are below the median. In this paper we concentrate more on determining the change in activity as measured by the water production rate over typically long periods of time of several apparitions.

## 2. Observations and Basic Analysis

The Solar Wind ANisotropies (SWAN) instrument on board the SOlar and Heliospheric Observatory (SOHO) satellite observes the whole sky in the emission of hydrogen Lyman-$\alpha$. The primary purpose of SWAN was to detect the Lyman-$\alpha$ emission of the atomic hydrogen atoms that pass through the solar system from interplanetary space (Bertaux et al. 1995). All-sky maps give a 3-D image of the effect of the solar flux on the loss of interstellar hydrogen that makes up the interplanetary background. Because SWAN is sensitive to H Ly$\alpha$ emission it also serves as a very useful detector for the fluorescence emission of the hydrogen comae of comets that is produced by the photodissociation of $H_2O$ and OH, which is typically the most abundant volatile constituent of comets (Bertaux et al. 1998). As such, SWAN has observed over 60 comets in the past 21 years from which water production rates have been calculated (Combi et al. 2019). SOHO is located in a halo orbit around the Earth-Sun L1 Lagrange point and so provides an excellent viewpoint for comets of sufficient brightness in the entire sky, whether in the northern or southern hemisphere, and not having any of the usual Earth horizon limitations of ground-based or low Earth orbit based observations. SWAN does have exclusion regions around the location of the Sun, as well as those obscured by the spacecraft itself in the general direction of the Earth as seen from SOHO. SWAN has been operated in an automatic mode for the last several years, providing daily scans of the entire sky with its 25 x 25 1-arc-second instrument field-of-view pixels. SWAN has two parts, one covering (essentially) the north heliographic hemisphere and the other covering the south. Images of comets are identified using their orbital elements.

Water production rates were determined from our time-resolved-model (TRM), as described by Mäkinen and Combi (2005). The TRM combines methods from Festou's (1981) vectorial model, the syndyname model of Keller and Meier (1976), and the Monte Carlo particle trajectory model of Combi and Smyth (1988a, 1988b). The spatial distribution of H Ly$\alpha$ coma is typically fitted by the TRM in an 8-degree radius circular field of view. For most weak to moderate comets an 8-degree field of view allows the comet+IPM background to asymptotically approach the IPM level so it captures most of the detectable coma. Since the fit of the model profile to the data integrates over the whole coma profile, the 8-degree field of view is not a critical parameter. Field stars and regions of data dropouts are manually masked, and the model fits the comet's hydrogen distribution and fits and subtracts the underlying interplanetary hydrogen background. Depending on the level of interference from background stars, the solar elongation angle, the local brightness of the interplanetary background, and the dust-to-gas brightness ratio, comets with visual magnitudes brighter than magnitude 10-12 are usually detectable by SWAN at a level so that water production rates can be determined.

Water production rates are calculated for each image, and a sample image and model analysis is shown in Figure 1. Because of the filling time of the field of view by hydrogen atoms the production rates usually represent an average over the previous 2-3 days, depending on the geocentric distance. If a comet is bright and spatially extended enough the TRM can analyze the various locations in the hydrogen coma in all images simultaneously. This can deconvolve the temporal/spatial information inherent in the coma, accounting for the time to produce H atoms by the photodissociation chain of $H_2O$ and OH as well as the transit time of H atoms across the coma. From this, daily-average water production rates from the vicinity of the nucleus are calculated. See Combi et al. (2005, 2014, 2019) for examples of its use. Calculating daily-average values is only useful for brighter comets than these, so useful results for 41P, 45P, or 46P were not obtained. This is borne out by rather large error bars for the extracted daily-average values, which are comparable to or even larger than the actual retrieved values. It is also worth noting here that power-law fits to the single-image production rates are not significantly different than those that would be obtained from daily-average values because over an average of ~2-3 days the largest change in heliocentric distance is only ~1% and the fitted intercepts and slopes are not this accurate.

Expected total uncertainties in water production rates determined from SWAN images of the hydrogen coma resulting from a combination of calibration and model description and parameters are expected to be on the order of ~30%. The g-factors are calculated from the composite solar Lyα flux data taken from the LASP web site at the University of Colorado (http://lasp.colorado.edu/lisird/lya/). The value from the face of the Sun seen by the comet is taken from the nearest time accounting for the number of days of solar rotation between the Earth and comet locations. The shape of the solar Lyα line profile is taken from the observation by Lemaire et al. (1998).

## 3. Comet 46P/Wirtanen

SWAN observations of 46P were obtained during the apparitions of 1997, 2002, 2008, and 2018. A summary of all the apparitions is given in Table 1. The data from 1997 were analyzed previously by Bertaux et al. (1999). While the results are similar both the model analysis (Mäkinen and Combi 2005) and the SWAN absolute calibration have changed (Combi et al. 2011a; Quémerais et al. 2009), with the largest calibration changes in the years 1996-2001. As mentioned above, the comet was too close to the Sun in the sky during the 2013 apparition to obtain useful images. The observational circumstances, g-factors, single-image water production rates and formal 1-σ uncertainties resulting from noise in the data and fitting procedure for the 2008 and 2018 apparitions are given in Table 2. Comet 46P reached perihelion in 2008 on 2008 February 2.50, and in 2018 on 2018 December 12.94. The similar results for the 1997 and 2002 apparitions are given the PDS archive (Combi 2017).

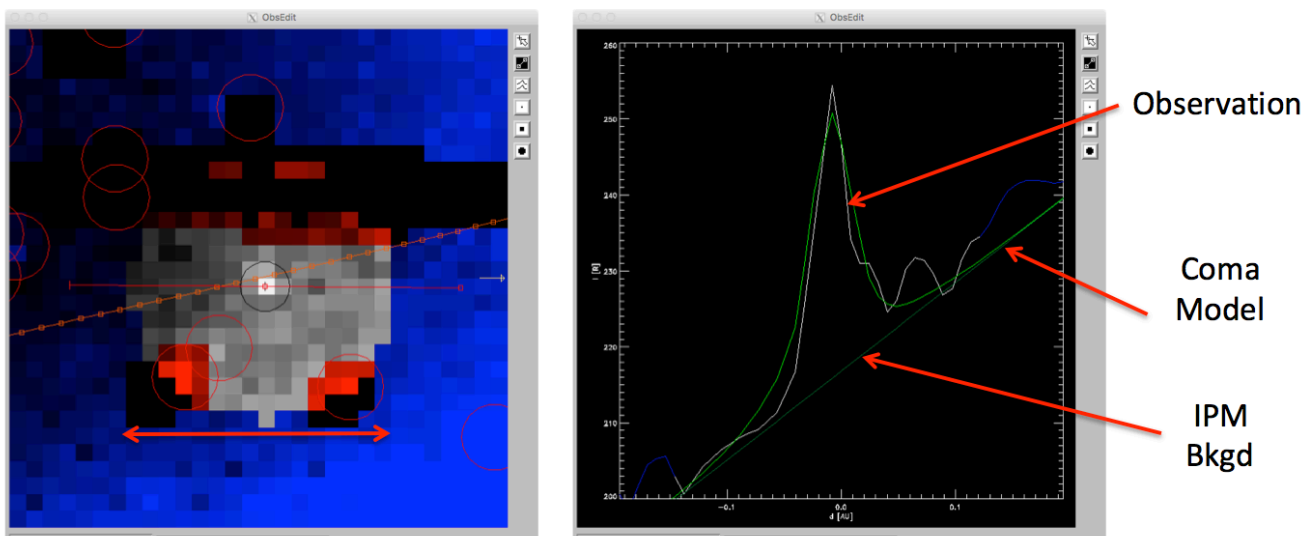

Figure 1. Sample of SWAN data and model analysis. On the left is a screen shot of a 30° x 30° region of the sky centered on comet 46P/Wirtanen in Lyman-α taken by the SWAN instrument on SOHO on 2008 February 4. The red arrow below the image on the left shows the 8° radius field of view included in the analysis of the comet emission. The thin red line through the middle of the comet shows the cut that corresponds to the profile in the right hand panel and the thin red line with the data points shows the path of the comet in the sky. The reddened regions show those areas not included because of field stars and data drop-outs. On the right is shown the brightness profile slice, indicated by the thin horizontal line in the image on the left showing the observation in white and the thicker green line showing the modeled comet profile, and the nearly straight thin green line below it showing the modeled interplanetary background (IPM).

Figure 2 shows the production rates plotted as a function of time in days from perihelion for all four apparitions. Figure 3 shows the water production rate in 46P/Wirtanen as a function of heliocentric distance with the two sets of power-law fits. Both the values all along the orbit decrease as do the slopes of the power laws. The values near perihelion drop by a factor of about 2 between the 1997/2002 combination and the 2008/2018 combination. The values farther from perihelion drop even more. Power-law fits of the water production rates with heliocentric distance seem to fall in two natural groups: one for 1997 and 2002, and the other for 2008 and 2018. In the 1997-2002 group the power-law fits were for pre-perihelion $Q = (1.68\pm0.17) \times 10^{28} r^{-3.6\pm0.7}$ and for post-perihelion $Q = (2.89\pm0.71) \times 10^{28} r^{-3.5\pm1.3}$ in $s^{-1}$. In the 2008-2018 group the power-law fits were for pre-perihelion $Q = (2.34\pm0.53) \times 10^{28} r^{-17.8\pm2.7}$ and for post-perihelion $Q = (1.90\pm0.16) \times 10^{28} r^{-8.6\pm0.7}$ also in $s^{-1}$.

Table 1

Summary of SOHO/SWAN Observations, Comet 46P/Wirtanen

| Apparition | q(AU) | # of Images | $r_H$ range (AU) |
|:---:|:---:|:---:|:---:|
| 1997 | 1.064 | 44 | 1.064 – 1.258 |
| 2002 | 1.058 | 28 | 1.062 – 1.208 |
| 2008 | 1.057 | 16 | 1.057 – 1.080 |
| 2013 | 1.052 | 0 | – |
| 2018 | 1.055 | 54 | 1.064 – 1.221 |

Notes to Table 1

$r_H$ = heliocentric distance (AU)

q = perihelion distance (AU)

Table 2

SOHO/SWAN Observations of Comet 46P/Wirtanen and Water Production Rates

| Date (UT) | r (AU) | Δ (AU) | g ($s^{-1}$) | Q ($s^{-1}$) | δQ ($s^{-1}$) |
|:---:|:---:|:---:|:---:|:---:|:---:|
| 2008 | | | | | |
| Jan 27.837 | 1.060 | 0.955 | 0.001656 | 9.01E+27 | 4.12E+27 |
| Jan 28.837 | 1.059 | 0.953 | 0.001656 | 9.48E+27 | 2.50E+27 |
| Jan 29.837 | 1.059 | 0.950 | 0.001657 | 1.15E+28 | 1.94E+27 |
| Jan 30.837 | 1.058 | 0.948 | 0.001655 | 1.16E+28 | 2.05E+27 |
| Feb 0.837 | 1.058 | 0.946 | 0.001656 | 1.15E+28 | 1.99E+27 |
| Feb 1.837 | 1.058 | 0.944 | 0.001657 | 1.44E+28 | 1.54E+27 |
| Feb 2.837 | 1.057 | 0.942 | 0.001658 | 1.54E+28 | 1.05E+27 |
| Feb 3.837 | 1.058 | 0.941 | 0.001658 | 1.66E+28 | 1.06E+27 |
| Feb 4.837 | 1.058 | 0.939 | 0.001659 | 1.67E+28 | 9.26E+26 |
| Feb 9.828 | 1.062 | 0.932 | 0.001659 | 1.64E+28 | 9.56E+26 |
| Feb 10.828 | 1.064 | 0.931 | 0.001660 | 1.46E+28 | 1.20E+27 |
| Feb 11.828 | 1.065 | 0.931 | 0.001660 | 1.03E+28 | 1.75E+27 |

| Date | | | | | |
|---|---|---|---|---|---|
| Feb 12.828 | 1.067 | 0.930 | 0.001660 | 1.04E+28 | 1.74E+27 |
| Feb 13.828 | 1.069 | 0.930 | 0.001661 | 1.01E+28 | 1.62E+27 |
| Feb 14.828 | 1.071 | 0.929 | 0.001662 | 1.14E+28 | 1.51E+27 |
| Feb 18.828 | 1.080 | 0.929 | 0.001664 | 1.10E+28 | 1.90E+27 |
| 2018 | | | | | |
| Nov 16.068 | 1.116 | 0.208 | 0.001404 | 3.65E+27 | 8.37E+26 |
| Nov 17.068 | 1.112 | 0.203 | 0.001399 | 2.47E+27 | 9.61E+26 |
| Nov 18.068 | 1.108 | 0.198 | 0.001401 | 5.38E+27 | 5.92E+26 |
| Nov 19.068 | 1.104 | 0.193 | 0.001399 | 4.19E+27 | 9.80E+26 |
| Nov 20.068 | 1.100 | 0.189 | 0.001398 | 5.96E+27 | 5.16E+26 |
| Nov 21.068 | 1.096 | 0.184 | 0.001400 | 3.63E+27 | 9.09E+26 |
| Nov 22.068 | 1.092 | 0.179 | 0.001391 | 3.56E+27 | 1.11E+27 |
| Nov 23.046 | 1.089 | 0.174 | 0.001393 | 6.16E+27 | 5.32E+26 |
| Nov 24.047 | 1.086 | 0.169 | 0.001392 | 6.49E+27 | 6.55E+26 |
| Nov 25.046 | 1.083 | 0.165 | 0.001386 | 4.42E+27 | 1.67E+27 |
| Nov 27.039 | 1.077 | 0.155 | 0.001368 | 7.05E+27 | 5.29E+26 |
| Nov 28.039 | 1.075 | 0.151 | 0.001370 | 5.62E+27 | 7.12E+26 |
| Dec 2.011 | 1.066 | 0.134 | 0.001380 | 5.52E+27 | 2.85E+26 |
| Dec 3.010 | 1.064 | 0.130 | 0.001382 | 7.08E+27 | 6.90E+26 |
| Dec 4.011 | 1.062 | 0.126 | 0.001377 | 7.70E+27 | 6.64E+26 |
| Dec 4.989 | 1.061 | 0.122 | 0.001376 | 6.38E+27 | 7.25E+26 |
| Dec 5.982 | 1.060 | 0.118 | 0.001378 | 5.96E+27 | 5.48E+26 |
| Dec 6.982 | 1.058 | 0.115 | 0.001374 | 6.22E+27 | 4.58E+26 |
| Dec 7.960 | 1.058 | 0.112 | 0.001381 | 7.50E+27 | 1.73E+27 |
| Dec 9.932 | 1.056 | 0.106 | 0.001385 | 8.97E+27 | 7.98E+26 |
| Dec 10.925 | 1.056 | 0.103 | 0.001389 | 7.79E+27 | 3.67E+26 |
| Dec 22.976 | 1.064 | 0.101 | 0.001371 | 1.61E+28 | 2.17E+26 |
| Dec 25.005 | 1.068 | 0.106 | 0.001376 | 1.13E+28 | 2.04E+26 |
| Dec 26.005 | 1.070 | 0.108 | 0.001375 | 8.80E+27 | 2.76E+26 |
| Dec 27.030 | 1.073 | 0.112 | 0.001373 | 7.62E+27 | 3.30E+26 |
| Dec 28.033 | 1.075 | 0.115 | 0.001362 | 8.47E+27 | 2.80E+26 |
| Dec 29.033 | 1.078 | 0.119 | 0.001359 | 8.70E+27 | 2.72E+26 |
| Dec 30.059 | 1.081 | 0.123 | 0.001356 | 7.45E+27 | 4.43E+26 |
| Dec 31.059 | 1.084 | 0.127 | 0.001360 | 7.85E+27 | 4.42E+26 |
| 2019 | | | | | |
| Jan 1.061 | 1.087 | 0.132 | 0.001371 | 8.32E+27 | 2.88E+26 |
| Jan 2.062 | 1.090 | 0.136 | 0.001361 | 7.53E+27 | 4.11E+26 |
| Jan 3.061 | 1.093 | 0.141 | 0.001358 | 7.36E+27 | 6.99E+26 |
| Jan 5.089 | 1.101 | 0.151 | 0.001354 | 8.00E+27 | 4.25E+26 |

| Date (UT) | r | Δ | g | Q | δQ |
|---|---|---|---|---|---|
| Jan 6.089 | 1.105 | 0.156 | 0.001358 | 6.67E+27 | 3.36E+26 |
| Jan 8.089 | 1.113 | 0.167 | 0.001341 | 6.10E+27 | 5.42E+26 |
| Jan 9.09 | 1.117 | 0.172 | 0.001350 | 6.22E+27 | 4.75E+26 |
| Jan 11.09 | 1.126 | 0.184 | 0.001360 | 6.59E+27 | 4.72E+26 |
| Jan 12.09 | 1.131 | 0.189 | 0.001358 | 6.84E+27 | 4.42E+26 |
| Jan 13.09 | 1.136 | 0.195 | 0.001363 | 6.49E+27 | 4.89E+26 |
| Jan 14.09 | 1.141 | 0.201 | 0.001374 | 5.43E+27 | 5.59E+26 |
| Jan 15.09 | 1.146 | 0.207 | 0.001375 | 5.46E+27 | 6.12E+26 |
| Jan 16.09 | 1.151 | 0.213 | 0.001379 | 4.35E+27 | 7.20E+26 |
| Jan 17.09 | 1.156 | 0.220 | 0.001390 | 5.40E+27 | 6.27E+26 |
| Jan 18.09 | 1.161 | 0.226 | 0.001391 | 5.09E+27 | 6.49E+26 |
| Jan 19.09 | 1.167 | 0.232 | 0.001389 | 4.78E+27 | 7.54E+26 |
| Jan 20.09 | 1.173 | 0.239 | 0.001386 | 4.72E+27 | 7.51E+26 |
| Jan 21.09 | 1.178 | 0.245 | 0.001381 | 4.73E+27 | 8.19E+26 |
| Jan 22.09 | 1.184 | 0.252 | 0.001386 | 4.82E+27 | 8.75E+26 |
| Jan 23.09 | 1.190 | 0.259 | 0.001374 | 5.55E+27 | 7.93E+26 |
| Jan 24.091 | 1.196 | 0.265 | 0.001374 | 4.91E+27 | 9.61E+26 |
| Jan 25.091 | 1.202 | 0.272 | 0.001369 | 3.65E+27 | 1.52E+27 |
| Jan 26.091 | 1.209 | 0.279 | 0.001369 | 3.14E+27 | 1.58E+27 |
| Jan 28.091 | 1.221 | 0.293 | 0.001363 | 5.36E+27 | 1.21E+27 |

Notes. Date (UT)

r : Heliocentric distance (AU)

Δ: Geocentric distance (AU)

g: Solar Lyman-α g-factor (photons s$^{-1}$) at 1 AU

Q: Water production rates for each image (s$^{-1}$)

δQ: internal 1-sigma uncertainties

In 2002 there were two large outbursts of 46P during the outbound (post-perihelion) part of the orbit, 15 and 35 days after perihelion. These outbursts were also clearly seen in the visual magnitude variations (Yoshida 2002). The variation during the entire 2002 apparition was more irregular than the other apparitions observed, but the water production rate levels were still clearly larger than in 2008 and 2018.

It is worth seeing what the effect of the 2002 post-perihelion outbursts is on the power-law fit to the production rate variation with heliocentric distance. Eliminating the two outbursts from the power-law fit results in a new expression, Q = (1.91±0.31) x $10^{28}$ $r^{-2.2±0.8}$. This makes the post-perihelion power-law fit for the earlier time period somewhat flatter, but does not change the fact that the change over time still remains. It is also worth mentioning that before, between and after the two outbursts, the production rate levels of 1997 and 2002 are nearly the same. The general behavior is strikingly similar to the change in both overall activity level and power-law exponents seen in hyperactive short-period comet 103P/Hartley 2 (Combi et al. 2011a, 2011b) over a similarly long time period. The results for 103P are shown for comparison in Figure 4.

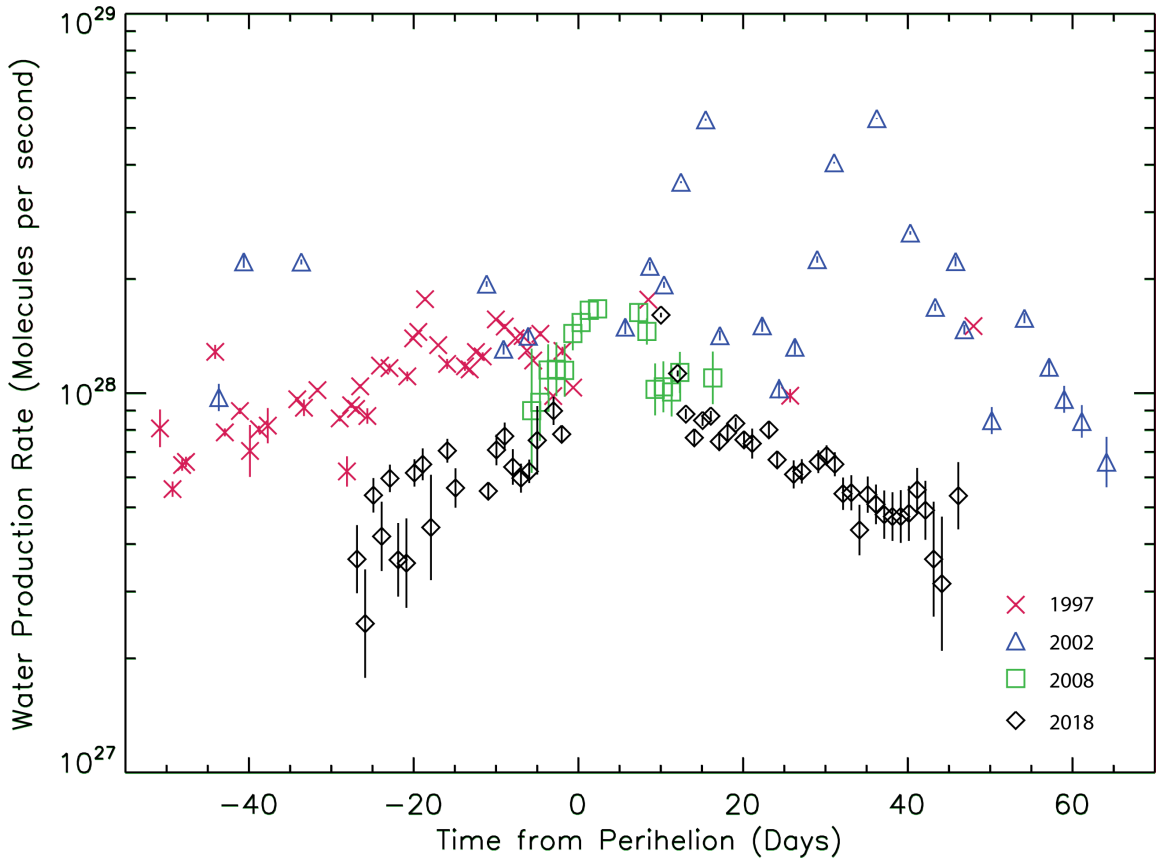

Figure 2. Single-image water production rates in comet 46P/Wirtanen are plotted as a function of time from perihelion in days. The error bars correspond to the respective 1σ formal uncertainties. The red x's show the values from the 1997 apparition, the blue triangles from 2002, the green squares from 2008, and the black diamonds from 2018. The vertical lines show the formal model fitting uncertainties.

4. Comet 45P/Honda-Mrkos-Pajdusakova

Comet 45P was observed by the SOHO SWAN H Lyα all-sky camera during the 2001, 2011, and 2017 apparitions. A summary of all the apparitions is given in Table 3. The production rate results, observational circumstances, and ancillary data for comet 45P are contained in and available from the PDS Small Bodies Node (Combi 2017). Figure 5 shows a plot of the production rates from all 3 apparitions plotted as a function of time from each perihelion in days. As can be seen in Figure 5, there is little consistent decrease in production rate over the 3 apparitions and 16 years, so data from all apparitions are taken together for the power-law fits. The power-law fit of water production rate as a function of heliocentric distance for the pre- and post-perihelion halves of the apparitions together was given by Combi et al. (2019) and are $Q = (8.6 \pm 1.6) \times 10^{26} \, r^{-5.9 \pm 0.3}$ and $Q = (6.3 \pm 0.9) \times 10^{27} \, r^{-3.7 \pm 0.2}$ in $s^{-1}$, respectively. Unlike comet 46P, 45P is not in the hyperactive class. The nucleus has a radius of 0.39 km (see Combi et al. 2019 for original references) and the water production rate near 1 AU is only a few times $10^{27} \, s^{-1}$. Production rates are much larger when the comet is near it rather small perihelion distance of ~0.5 au. Therefore, unlike the hyperactive comets, we suspect that the activity of 45P is controlled as a "normal" comet by sublimation of water ice rather than something more volatile like $CO_2$ and that the water in the coma mostly originates directly from the nucleus.

As can be seen in both Figure 5 and in the power-law fits, the activity is decidedly larger after perihelion than before, and so this is likely driven by a typical seasonal effect.

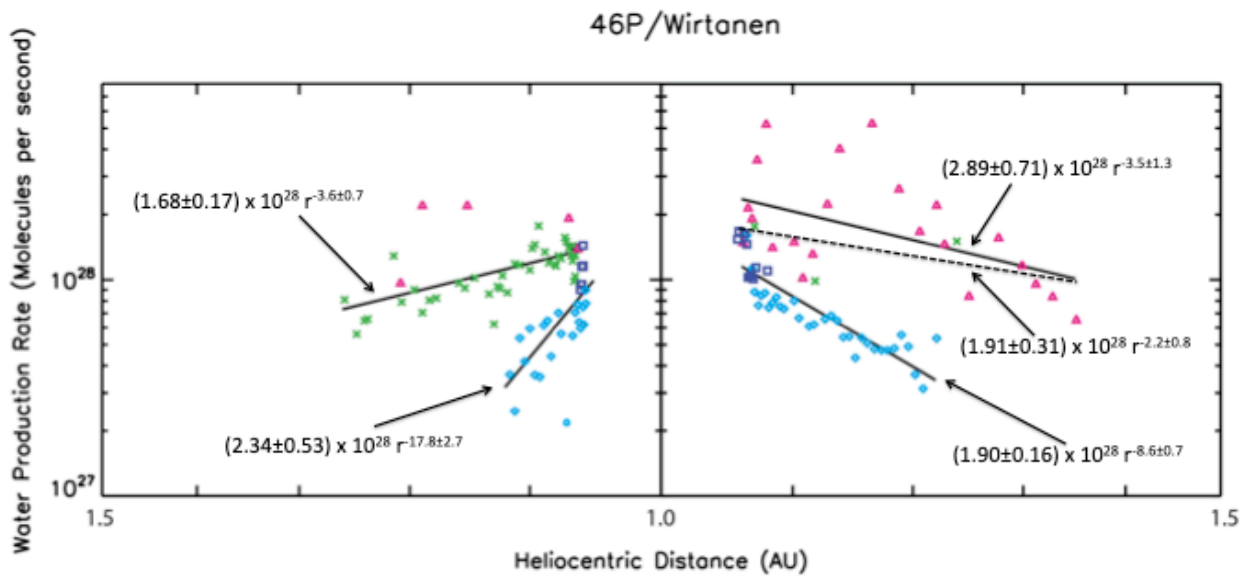

Figure 3. Single-image water production rates in comet 46P/Wirtanen are plotted as a function of heliocentric distance in AU. The pre-perihelion data are in the left half and the post-perihelion data in the right. Note that the perihelion distance is 1.06 AU and is responsible for the gap. The upper and lower straight lines represent the power-law fits to the 1997/2002 group and 2008/2018 group, respectively. The best-fit coefficients for the power-law fits are given in the text. The x's (green) show the values from the 1997 apparition, the triangles (magenta) from 2002, the squares (dark blue) from 2008, and the diamonds (cyan) from 2018.

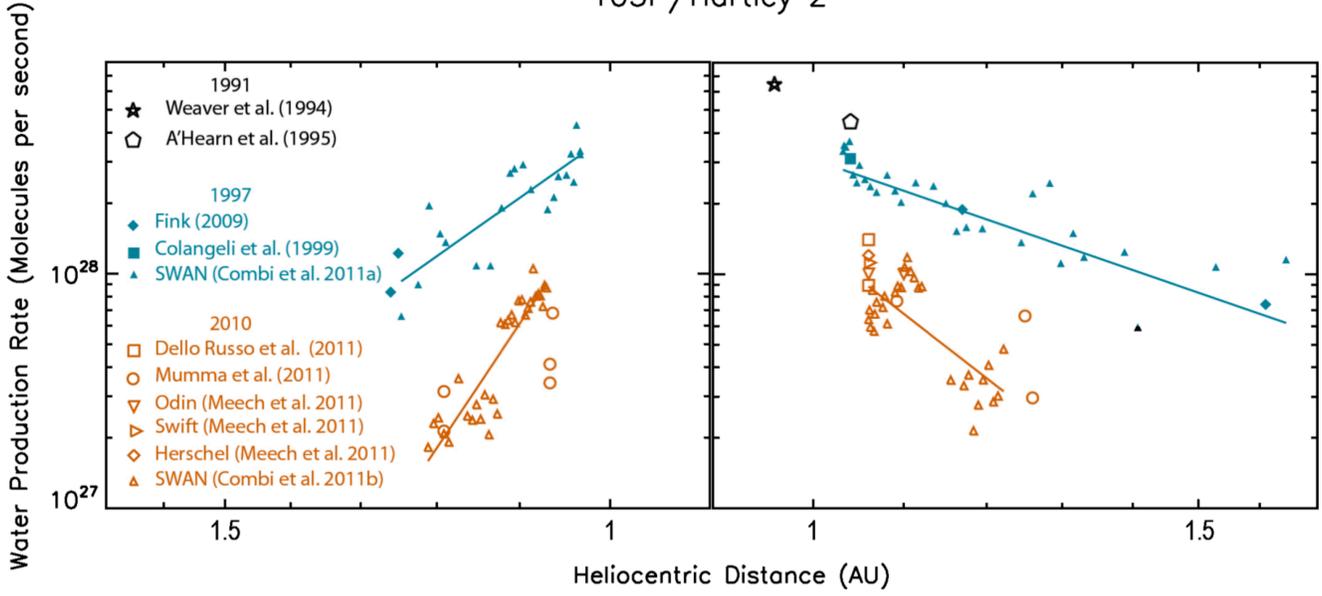

Figure 4. Water production rates in comet 103P/Hartley 2 in 1991 (black), 1997 (cyan), and 2010 (orange), from various sources and methods. The left panel gives the pre-perihelion results and the right panel gives the post-perihelion results. The power-law fits are the straight lines given for the 1997 SWAN data (above) and 2010 SWAN data (below) in $s^{-1}$ and are $3.94 \times 10^{28} \times r^{-6.6}$ and $3.08 \times 10^{28} r^{-3.2}$ for pre- and post-perihelion, respectively, in 1997 and $2.31 \times 10^{28} r^{-14.0}$ and $1.35 \times 10^{28} r^{-7.2}$ for pre- and post-perihelion, respectively, in 2010.

Table 3

Summary of SOHO/SWAN Observations, Comet 45P/Honda-Mrkos-Pajdusakova

| Apparition | q(AU) | # of Images | $r_H$ range (AU) |
| --- | --- | --- | --- |
| 2001 | 0.528 | 21 | 0.532 — 1.427 |
| 2006 | 0.530 | 0 | – |
| 2011 | 0.530 | 46 | 0.540 — 1.039 |
| 2017 | 0.532 | 26 | 0.535 – 1.058 |

Notes to Table 3

$r_H$ heliocentric distance (AU)

q – perihelion distance (AU)

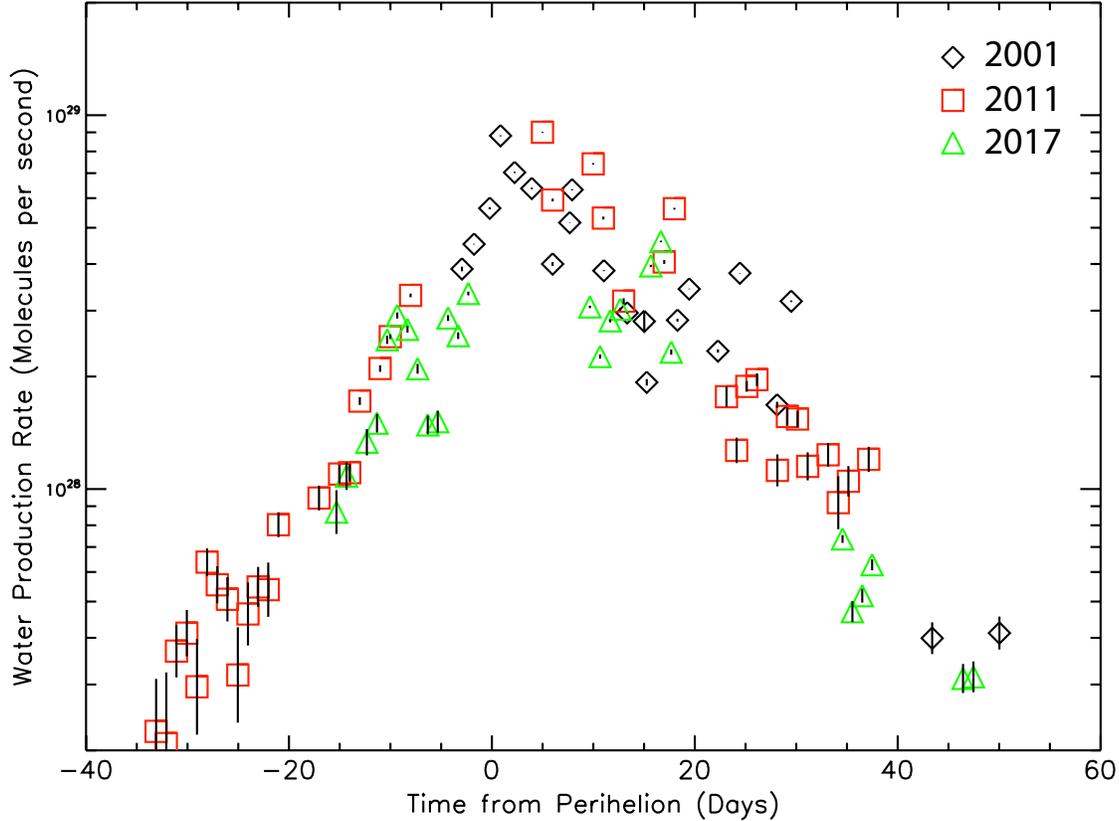

Figure 5. Single-image water production rates in comet 45P/Honda-Mrkos-Pajdusakova are plotted as a function of time from perihelion in days. The error bars correspond to the respective 1σ formal uncertainties. The black diamonds show the values from the 2001 apparition, the red squares from 2011, and the green triangles from 2017. The vertical lines are the formal model-fitting uncertainties.

## 5. Comet 41P/Tuttle-Giacobini-Kresak

Comet 41P was observed by the SOHO SWAN H Lyα all-sky camera during the 2001, 2006, and 2017 apparitions. The production rate results and ancillary data for comet 41P are contained in and available from the PDS Small Bodies Node (Combi 2017). A summary of the data from the multiple apparitions is given in Table 4. As can be seen by an examination of Figure 6 that shows the water production rates from all three apparitions plotted as functions of time from perihelion in days, the comet activity has changed markedly over the 16 years. In 2001 there appear to have been two outbursts or extended brightenings reaching peaks 35 days and 18 days before perihelion. In 2006 there was a similar brightening reaching a peak at about 25 days before perihelion. As discussed above, comet 41P was known for outbursts over the previous almost 50 years. On the other hand, the 2017 apparition was rather flat and uneventful, with production rates remaining close to the lower end of SWAN detectability with values around 3 x $10^{27}$ $s^{-1}$. Levels in 2017 were factors of 2-4 below comparable times in 2006 and factors of 10-30 below those in 2001.

Table 4

Summary of SOHO/SWAN observations comet 41P/Tuttle-Giacobini-Kresak

| Apparition | q(AU) | # of Images | $r_H$ range (AU) |
|---|---|---|---|
| 2001 | 1.052 | 34 | 1.052 — 1.227 |
| 2006 | 1.048 | 27 | 1.050 - 1.132 |
| 2011 | 1.049 | 0 | - |
| 2017 | 1.045 | 13 | 1.045 — 1.052 |

$r_H$ heliocentric distance (AU)

q - perihelion distance (AU)

In the SWAN 61-comet survey (Combi et al. 2019), power-laws of water production rate versus heliocentric distance were fitted to three different pre- or post-perihelion set of individual apparition data sets, however, most did not lend themselves to being represented by a power law. Certainly, using all sets of apparition data together was not appropriate for a power-law representation with heliocentric distance.

## 6. Summary

Comet 46P/Wirtanen was observed by the SOHO SWAN H Lyα all-sky camera during the 1997, 2002, 2008, and 2018 apparitions. Water production rates were determined from each of the images using our standard model analysis. We find a significant change between the 1997/2002 and 2008/2018 apparitions with a marked decrease in overall production rates throughout the apparitions as well as a large steepening of the variation of water production rate with heliocentric distance. The changes are highly reminiscent of those that occurred in comet 103P/Hartley 2, another so-called hyperactive comet, between the 1997 and 2011 apparitions (Combi et al. 2011a).

It remains to be seen in observations from the next apparitions of both 103P and 46P whether the decreases in activity and changes in heliocentric distance dependencies are very long-term trends of fading away or if these hyperactive comets go through cycles of decreasing and increasing activity. This will be answered for 103P during the next apparition in 2023, which is very favorable for Earth- and near-Earth-based observation. Unfortunately the 2024 apparition geometry for 46P is less favorable. Another question is whether in 46P, like the other hyperactive comets 73P/Schwassmann-Wachmann 1 (Fougere et al. 2012) and 103P (Fougere et al. 2013), a major part of the water production (and possible other volatiles) is from an extended source of icy grains/chunks rather than from direct sublimation of the nucleus. Potentially, close-up observations enabled by the favorable observing geometry of 46P in 2018/2019 could answer this question.

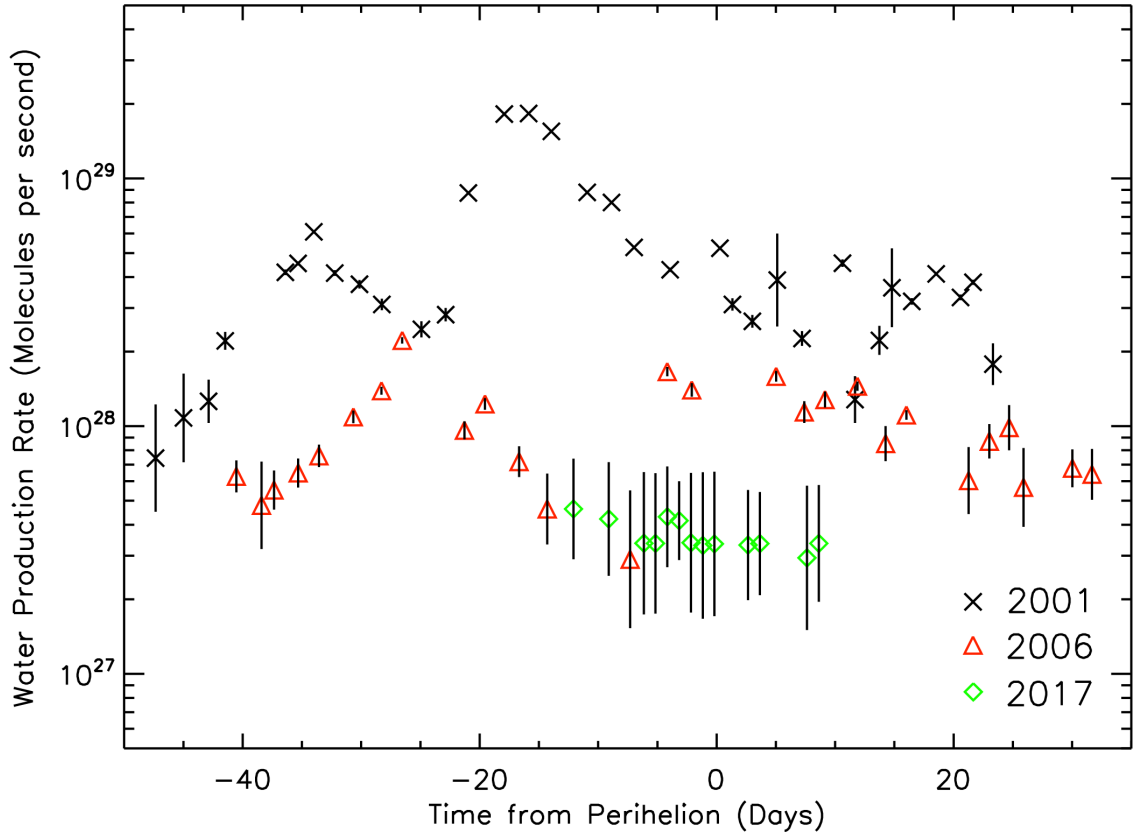

Figure 6. Single-image water production rates in comet 41P/Tuttle-Giacobini-Kresak. are plotted as a function of time from perihelion in days. The error bars correspond to the respective 1σ formal uncertainties. The black x's show the values from the 2001 apparition, the red triangles from 2006, and the green diamonds from 2017. The vertical lines are the formal model-fitting uncertainties.

Comet 45P/Honda-Mrkos-Pajdusakova was observed by SOHO SWAN during the 2001, 2011, and 2017 apparitions. Although 45P has the smallest perihelion distance of the three comets described here and its nucleus is rather small, there was little decrease in overall activity as measured by the water production rate over four apparitions. Note again that the 2006 apparition had poor observing geometry from near the Earth. Having a rather steep slope with heliocentric distance, a moderate asymmetry about perihelion, and a rather consistent level of activity over 17 years, it is rather more similar to the behavior of comet 67P/Churyumov-Gerasimenko (Bertaux et al. 2014) than it is to 46P or 41P, which have faded quite dramatically over a similar length of time and are more similar to the hyperactive comet 103P/Hartley 2.

Comet 41P/Tuttle-Giacobini-Kresak was observed by SOHO SWAN during the 2001, 2006, and 2017 apparitions. In the case of 41P, the observing geometry for Earth-based or near-Earth-based observers was poor for the 2011 apparition. The activity of Comet 41P, which has been noted for a number of large outbursts over the last nearly 50 years, has decreased markedly from 2001 to 2017. Two outbursts were seen in 2001 and one in 2006, but the decrease in water production was on the order of a factor of 3 or more from 2006 to 2017 and for comparable times along the orbit the decrease was a factor of 10—30 from 2001 to 2017.

Acknowledgements: SOHO is an international mission between ESA and NASA. M. Combi acknowledges support from NASA grant 80NSSC18K1005 from the Solar System Observations Program. T.T. Mäkinen was supported by the Finnish Meteorological Institute (FMI). J.-L. Bertaux and E. Quémerais acknowledge support from CNRS and CNES. We obtained orbital elements from the JPL Horizons web site (http://ssd.jpl.nasa.gov/horizons.cgi). The composite solar Lyα data was taken from the LASP web site at the University of Colorado (http://lasp.colorado.edu/lisird/lya/). We acknowledge the personnel that have been keeping SOHO and SWAN operational for over 20 years, in particular Dr. Walter Schmidt at FMI. We also acknowledge the support of R. Coronel by the Undergraduate Research Opportunity Program of the University of Michigan. We also thank the two reviewers for their careful reading and helpful suggestions that have improved the paper.